\documentclass{PoS}
\usepackage{braket}
\usepackage[noadjust]{cite}
\title{Rare Leptonic B decays}

\ShortTitle{Rare Leptonic B decays}

\author{\speaker{Gilberto Tetlalmatzi-Xolocotzi}\\
        Nikhef, Science Park 105, NL-1098 XG Amsterdam, Netherlands\\
        E-mail: \email{gtx@nikhef.nl}}

\abstract{The transitions $B_{s,d}\rightarrow \ell^+\ell^-$ (for $\ell=e, \mu, \tau$), although extremely suppressed within the SM, are exceptionally clean channels particularly sensitive to New Physics contributions. So far only the branching ratio for the process  $B_{s}\rightarrow \mu^+\mu^-$ has been measured experimentally and, although the results are in agreement with the SM prediction, there is still room for NP effects. Here we provide an overview of the theoretical status of rare decays. We present strategies that can allow us to discriminate between different NP scenarios. Finally,  we include a discussion of the effects of nontrivial new CP violating phases on the observables associated with $B_{s}\rightarrow \mu^+\mu^-$.}

\FullConference{The International Conference on B-Physics at Frontier Machines - BEAUTY2018\\
		6-11 May, 2018\\
		La Biodola, Elba Island, Italy}

\begin{document}

\section{Introduction}

\begin{figure}\label{fig:status}	
	%\vspace{1.6cm}
	\includegraphics[width=0.5\textwidth]{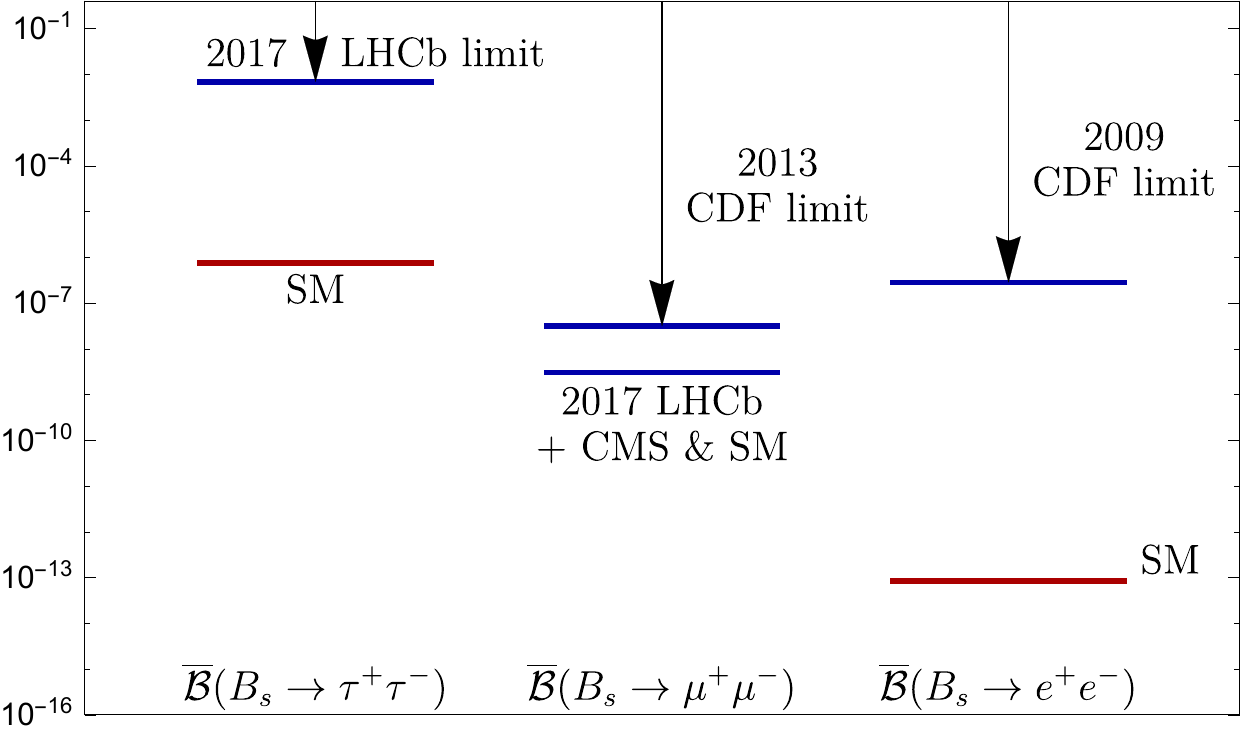}
	\hspace{0.25cm}
	\includegraphics[width=0.5\textwidth]{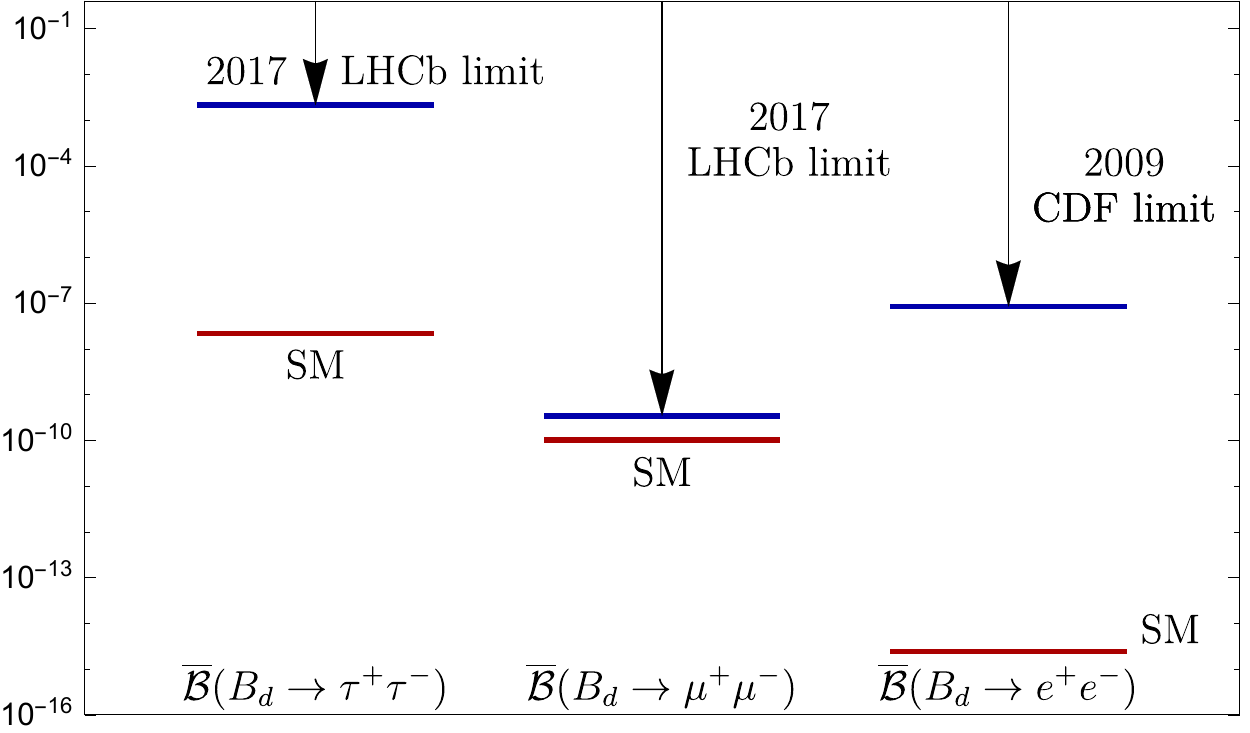}
	%\hspace{0.005cm}
	%\includegraphics[width=0.5\textwidth]{fig1b.pdf}
	\caption{Current experimental and theoretical status of the different rare $B$ meson decays.}
\end{figure}

In the Standard Model (SM) the transitions $B_{s,d}\rightarrow \ell^+\ell^-$ are the result of purely quantum mechanical processes, i.e. they are the result of the interchange of virtual particles at the loop level. Moreover the associated decay probabilities are proportional to the square of the mass of the lepton in the final state. Therefore, for electrons and muons ($\ell=e,~\mu$) they turn out to be extremely small (helicity suppression). For  $\ell=\tau$, the helicity suppression is not very effective due to the relatively big value of $m_{\tau}$, however $\tau$ leptons are difficult to be reconstructed experimentally. Due to these features the processes $B_{s,d}\rightarrow \ell^+\ell^-$ receive the generic name of ``rare  $B$ decays''.
They  have unique properties that make them particularly attractive, for instance non perturbative contributions are well under control. Moreover, they are particularly sensitive to New Physics (NP) contributions from scalar and pseudoscalar particles \cite{Bsmumu-ADG, BFGK, Altmannshofer:2017wqy}. The current experimental and theoretical status of the different rare decays is summarized in Fig. \ref{fig:status}. At the moment only $B_{s}\rightarrow \mu^+\mu^-$ has been measured experimentally, the combination of the LHCb and CMS determinations yields \cite{LHCb-2017,CMSmumu}
\begin{eqnarray}\label{eq:Exp}
\overline{\mathcal{B}}(B_s\rightarrow \mu^+ \mu^-)|_{\rm LHCb'17 + CMS'13}&=&(3.00\pm 0.5)\times 10^{-9},
\end{eqnarray}
\noindent
in good agreement with the SM prediction (the usage of the  ``$\overline{\mathcal{B}}$'' notation will be explained later). There is also a determination by ATLAS from 2016 that shows compatibility
with the SM at the $2~\sigma$ level and can be found in \cite{Aaboud:2016ire}. \\
\noindent
\\
Here we will give an overview of the theory behind  leptonic rare $B$ decays (for the study of NP in semileptonic decays see for example \cite{Feruglio:2018jnu, Mahmoudi, Ciuchini,  Altmannshofer:2015sma}). In addition to the branching ratio, we present extra observables ($\tau^s_{\ell\ell}$, ${\cal A}_{\Delta\Gamma_s}^{ \ell\ell}$, ${\cal C}_{\ell\ell}$ and ${\cal S}_{\ell\ell}$) that give us the power to unveil potential NP effects and to discriminate among different models. In view of the current experimental information available we show that NP effects are allowed. Furthermore, we describe how having NP short distance contributions independent of the  flavour of the lepton in the final state can lead to enhancements on the decay channels $\overline{\mathcal{B}}(B_{s,d}\rightarrow e^+ e^-)$  making them experimentally accessible while keeping $\overline{\mathcal{B}}(B_{s,d}\rightarrow \tau^+ \tau^-)$ as in the SM.  In the last part, we comment on the possibility of using $B$ meson rare decays for pinning down NP phases.

\pagebreak
\section{Theoretical Formalism}

The Hamiltonian to describe $\bar{B}^0_s\rightarrow \ell^+\ell^-$ and $B^0_s\rightarrow \ell^+\ell^-$  transitions is

\begin{equation}\label{Heff}
{\cal H}_{\rm eff}=-\frac{G_{\rm F}}{\sqrt{2}\pi} V_{ts}^\ast V_{tb} \alpha
\bigl[C^{\ell\ell}_{10} O_{10} + C^{\ell\ell}_P O_P +   C^{\ell\ell}_{S} O_S + 
C_{10}^{\ell\ell'} O_{10}' + C_{S}^{\ell\ell'} O_S' + C_P^{\ell\ell'} O_P' \bigr] + \rm{h.c.}, \nonumber
\end{equation}
\noindent
where the heavy degrees of freedom have been integrated out and are described by the Wilson coefficients $C^{\ell\ell}_{10, S, P}$ and $C_{10, S, P}^{\ell\ell'}$. 
In the SM $C_{10}$ is the only non-vanishing Wilson Coefficient and turns out to be real whereas $C^{\ell\ell}_{S, P}=C_{10, S, P}^{\ell\ell'}=0$. The explicit expressions for the operators in Eq.~(\ref{Heff}) are

\begin{eqnarray}\label{operators}
O_{10}  = (\bar s \gamma_\mu P_L b) (\bar\ell\gamma^\mu \gamma_5\ell),\quad
O_P     = m_b (\bar s P_R b)(\bar \ell \gamma_5 \ell),\quad
O_S=m_b (\bar s P_R b)(\bar \ell \ell)\nonumber\\
O'_{10}  = (\bar s \gamma_\mu P_L b) (\bar\ell\gamma^\mu \gamma_5\ell),\quad
O'_P     = m_b (\bar s P_R b)(\bar \ell \gamma_5 \ell),\quad
O'_S=m_b (\bar s P_R b)(\bar \ell \ell)
\end{eqnarray}
with $P_{L/R}\equiv\frac{1}{2}\left(1 \mp \gamma_5\right)$.\\

\noindent
For future convenience, we introduce the scalar and pseudoscalar functions $P^s_{\ell\ell}$ and $S^s_{\ell\ell}$ connected with the Wilson coefficients introduced in Eq.~(\ref{Heff}) according to

\begin{eqnarray}\label{P-expr}
P^s_{\ell\ell}&\equiv& \frac{C^{\ell\ell}_{10}-C^{\ell\ell'}_{10}}{C_{10}^{\rm SM}} +\frac{M_{B_s}^2}{2  m_\ell}
\left(\frac{m_b}{m_b+m_s}\right)\left[\frac{C^{\ell\ell}_P-C^{\ell\ell'}_P}{C_{10}^{\rm SM}}\right]=|P^s_{\ell\ell}|e^{i\varphi_P},\nonumber\\
%\end{equation}
%\begin{equation}\label{S-expr}
S^s_{\ell\ell}&\equiv& \sqrt{1-4\frac{m_\ell^2}{M_{B_s}^2}}
\frac{M_{B_s} ^2}{2 
	m_\ell}\left(\frac{m_b}{m_b+m_s}\right)
\left[\frac{C^{\ell\ell}_S-C^{\ell\ell'}_S}{C_{10}^{\rm SM}}\right]=|S^s_{\ell\ell}|e^{i\varphi_S}.\nonumber
\end{eqnarray}
\noindent
Using the effective Hamiltonian in Eq.~(\ref{Heff}), the corresponding SM ``theoretical'' branching fraction can be computed leading to \cite{Buchalla:1995vs}

\begin{eqnarray}\label{eq:BrSMTHeso}
\mathcal{B}(B_s \to \ell^+\ell^-)|_{\rm SM }=
%\Biggl[
m^2_{\ell}
\frac{\tau_{B_s}G^4_F M^4_W \sin^4\theta_W}{8\pi^5}\Bigl|V_{ts}V^*_{tb}\Bigl|^2
f^2_{B_s}M_{B_s}\sqrt{1-\frac{m^2_{\ell}}{M^2_{B_s}}}\Bigl|C^{SM}_{10}\Bigl|^2,
\end{eqnarray}
\noindent
\noindent
where the non-perturbative hadronic effects are accounted for by the decay constant $f_{B_s}$ calculated through lattice techniques \cite{Aoki:2016frl, Carrasco:2013naa, Dowdall:2013tga, Christ:2014uea, Aoki:2014nga, Witzel:2013sla, Na:2012kp, McNeile:2011ng, Bazavov:2011aa, Gamiz:2009ku, Bernardoni:2014fva, Carrasco:2013zta, Carrasco:2013iba, Bernardoni:2012ti, Carrasco:2012de, Blossier:2011dk, Dimopoulos:2011gx, Blossier:2009hg} with a current precision of $\mathcal{O}(2\%)$. So far $B_s-\bar{B}_s$ mixing has been ignored; once this effect is taken into account, the dynamics of the rare decays becomes time dependent. To discuss in more detail this effect, we make a small digression to introduce some basic terminology in neutral $B$ mixing. The time evolution of the two state system  $| B_s (t) \rangle-| \bar{B}_s(t) \rangle$ is given by

\begin{equation}
i \frac{d}{dt} \left(
\begin{array}{c}
| B_s (t) \rangle
\\
| \bar{B}_s(t) \rangle
\end{array}
\right)
= \left(
\hat{M}^s - \frac{i}{2} \hat{\Gamma}^s
\right)
\left(
\begin{array}{c}
| B_s(t) \rangle
\\
| \bar{B}_s(t) \rangle
\end{array}
\right) \; ,
\end{equation}

\noindent
where $\hat{M}^s$ and $\hat{\Gamma}^s$ are $2\times 2$ matrices  that due to electroweak interactions are non-diagonal. The off-diagonal elements of  $\hat{M}^s$,
$M^s_{12}=M^{s*}_{21}$, receive contributions from virtual internal particles. On the other hand the off-diagonal entries of  $\hat{\Gamma}^s$, $\Gamma^s_{12}=\Gamma^{s*}_{21}$, receive contributions from on-shell particles only.
\noindent
After diagonalizing $\hat{M}^s - \frac{i}{2} \hat{\Gamma}^s$ the physical states 
$| B_H (t) \rangle$ and $| B_L (t) \rangle$ are found. The corresponding masses and decay rates are denoted by $M^s_H$, $M^s_L$ and  $\Gamma^s_H$, $\Gamma^s_L$ respectively. In this report we are interested in the following masses and decay rate differences 

\begin{eqnarray}
\Delta M_s=M^s_H - M^s_L  \quad \quad  \Delta \Gamma_s=\Gamma^s_H - \Gamma^s_L.
\end{eqnarray}
\noindent
We now continue with the our main line of discussion and summarize
the steps followed towards the determination of the branching ratio for the transitions $B_s\rightarrow \ell^+ \ell^-$ once the time dependence is considered. Since measuring the helicities of the final states is challenging to perform, we begin by adding over the possible final states helicities

\begin{eqnarray}
\Gamma(B_s^0(t)\to \ell^+\ell^-)\equiv \sum_{\lambda={\rm L,R}}
\Gamma(B_s^0(t)\to \ell^+_\lambda \ell^-_\lambda).\nonumber
\end{eqnarray}
\noindent
Moreover, $B^0_s$ and $\bar{B}^0_s$ tagging  is not easy to do in experiments, thus we consider the untagged rate

\begin{eqnarray}
\langle \Gamma(B_s(t)\to \ell^+\ell^-)\rangle=\Gamma(B^0_s(t)\to \ell^+\ell^-) + \Gamma(\bar B^0_s(t)\to \ell^+\ell^-).
\end{eqnarray}
\noindent
What is measured experimentally is the time integrated branching ratio \cite{Bsmumu-ADG}

\begin{eqnarray}\label{defBrExp}
\overline{{\mathcal B}}\left(B_s \to \ell^+\ell^-\right) 
&\equiv& \frac{1}{2}\int_0^\infty \langle \Gamma(B_s(t)\to \ell^+\ell^-)\rangle\, dt.
\end{eqnarray}

\noindent
In the remainder of this work we will refer to $\overline{{\mathcal B}}\left(B_s \to \ell^+\ell^-\right)$ as the ``experimental'' branching ratio.\\

\noindent
The connection between the ``experimental'' branching fraction and its SM ``theoretical'' counterpart $\mathcal{B}(B_s \to \ell^+\ell^-)|_{\rm SM}$ is
established using the effective Hamiltonian in Eq.~(\ref{Heff})  \cite{BFGK}

\begin{equation}\label{R-def}
\overline{{\mathcal B}}(B_s \to \ell^+\ell^-)
=
\mathcal{B}(B_s \to \ell^+\ell^-)|_{\rm SM} \times\Biggl\{
\left[\frac{1+y_s\cos(2\varphi_P -\phi^{NP}_s) }{1-y_s^2}  \right] |P^s_{\ell\ell}|^2+
\left[\frac{1-y_s\cos(2\varphi_S-\phi^{NP}_s)}{1-y_s^2}  \right] |S^s_{\ell\ell}|^2\Biggl\},\nonumber
\end{equation}

\noindent
where the parameter $y_s=\Delta \Gamma_s /\Gamma_s$ accounts for neutral $B$ mixing effects. \\

\noindent
The phase $\phi^{NP}_s$  quantifies  possible NP contributions to mixing and can be determined using experimental information from the transition $B^0_s\rightarrow J/\psi \phi$ and processes with similar dynamics leading to \cite{Amhis:2016xyh, DeBruyn:2014oga, Charles:2015gya}
\begin{eqnarray}
\phi^{NP}_s = (0.4 \pm 1.9)^{\circ}.
\end{eqnarray} 
\noindent
In the absence of NP effects, $P^s_{\ell\ell}=1$ and $S^s_{\ell\ell}=0$, hence the experimental branching fraction in Eq.~(\ref{defBrExp}) reduces to $\overline{{\mathcal B}}(B_s \to \ell^+\ell^-)|_{\rm SM }
=1/(1-y_s)\times \mathcal{B}(B_s \to \ell^+\ell^-)|_{\rm SM }$. Thus, even in the purely SM case, and as the result of neutral meson mixing, there is a mismatch between the experimental branching ratio introduced in Eq.~(\ref{R-def}) and the theoretical one in Eq.~(\ref{eq:BrSMTHeso}) given by the factor  $1/(1-y_s)$~~\cite{Bsmumu-ADG}.\\

\noindent
Using Eqs.~(\ref{eq:BrSMTHeso}) and (\ref{R-def})  we obtain \cite{Fleischer:2017ltw, Bobeth:2013uxa}
\begin{eqnarray}\label{eq:theo}
\overline{\mathcal{B}}(B_s\rightarrow \ell^+\ell^-)|_{\rm SM }=(3.57\pm 0.16)\times 10^{-9},
\end{eqnarray}
as discussed in \cite{Beneke:2017vpq} possible electromagnetic corrections below $m_b$ lead to modifications in Eq.~(\ref{eq:theo}) of $\mathcal{O}(1\%)$.\\ 

\noindent
To probe for NP effects we consider the ratio

\begin{equation}\label{Rellells}
\overline{R}_{\ell\ell}^s \equiv
\frac{\overline{\mathcal{B}}(B_s\to\ell^+\ell^-)}{\overline{\mathcal{B}}
	(B_s\to\ell^+\ell^-)|_{\rm SM}}\nonumber\\
= 
\left[\frac{1+y_s\cos(2\varphi^{\ell\ell}_{P_s}-\phi_s^{\rm NP})}{1+y_s} \right] |P_{\ell\ell}|^2 + 
\left[\frac{1-y_s\cos(2\varphi^{\ell\ell}_{S_s}-\phi_s^{\rm NP})}{1+y_s} \right] |S_{\ell\ell}|^2.
\end{equation}
\noindent
If we consider trivial CP violating phases $\varphi^{\ell\ell}_{P_s}, \varphi^{\ell\ell}_{S_s} \in \{0, 2\pi\} $ then $P^s_{\mu\mu}$ and $S^s_{\mu\mu}$ are real quantities.\\

\noindent
The ratio in Eq.~(\ref{Rellells}) becomes

\begin{eqnarray}
\overline{R}_{\ell\ell}^s\approx P^{s~2}_{\ell\ell}+S^{s~2}_{\ell\ell},
\end{eqnarray}

\noindent
and the geometrical region described by the observable is a  circle of radius $\sqrt{\overline{R}_{\ell\ell}^s}$ in the plane defined by $P^s_{\ell\ell}$ and $S^s_{\ell\ell}$. Using Eqns. (\ref{eq:Exp}) and (\ref{eq:theo}) we proceed with the corresponding numerical evaluation obtaining
\begin{eqnarray}\label{eq:Rres}
\overline{R}^s_{\mu\mu}\bigl|_{\hbox{\tiny LHCb'17+CMS}} = 0.84\pm 0.16.
\end{eqnarray}

\begin{figure}
	\begin{center}
		\label{fig:PvsSoAGF}
		%\vspace{1.6cm}
		\includegraphics[width=0.6\textwidth]{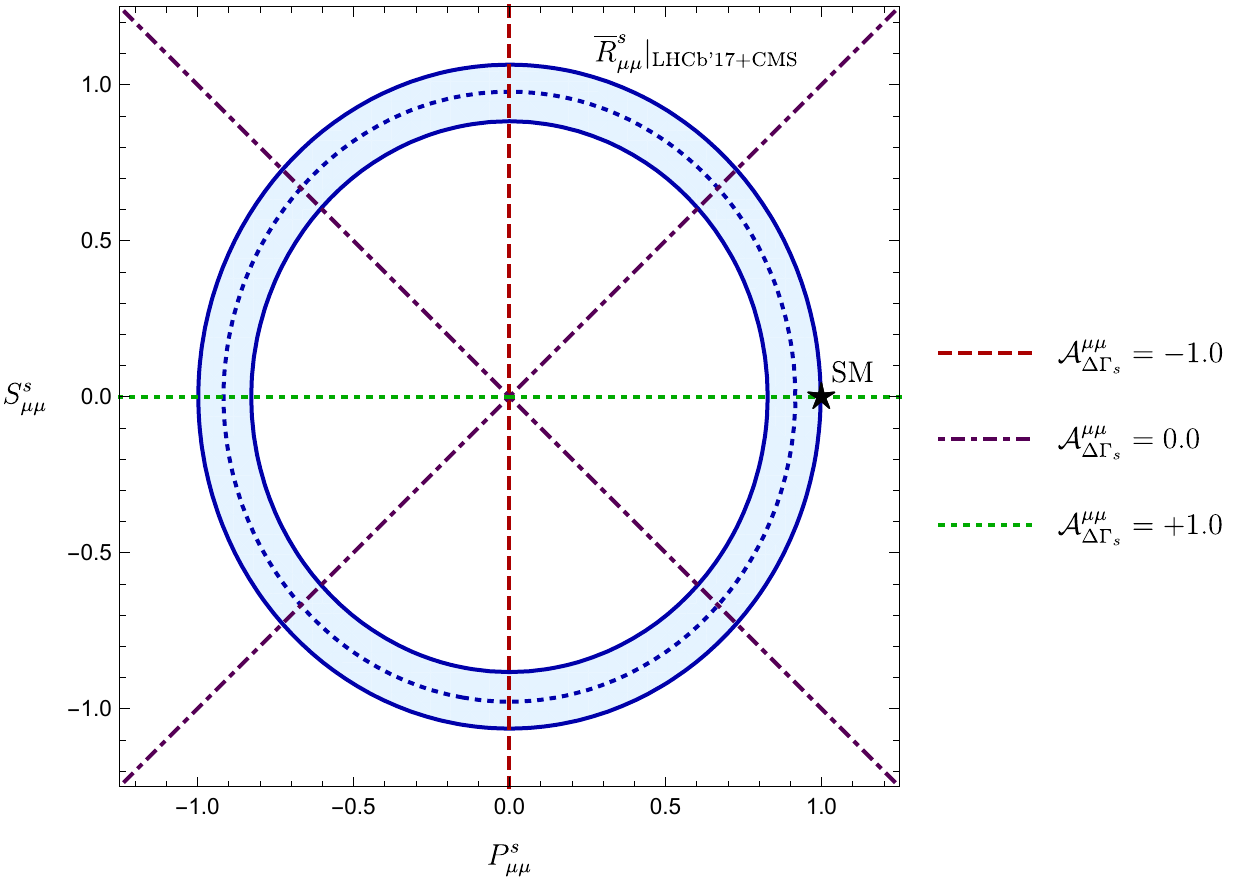}
		%\hspace{0.005cm}
		%\includegraphics[width=0.5\textwidth]{fig1b.pdf}
	\end{center}
	\caption{Allowed regions for $P^s_{\mu\mu}$ and $S^s_{\mu\mu}$ considering the current measurements of $\bar{R}^s_{\mu\mu}$ and $\mathcal{A}^{\mu\mu}_{\Delta \Gamma_s}$. The dashed straight lines correspond to hypothetical values for $\mathcal{A}^{\mu\mu}_{\Delta \Gamma_s}$.}
\end{figure}

\noindent
The allowed region corresponding to Eq.~(\ref{eq:Rres}) leads to the blue circular band in the $P^s_{\mu\mu}-S^s_{\mu\mu}$ plane  shown in Fig. \ref{fig:PvsSoAGF} \cite{Fleischer:2017ltw}. We can see how the observable $\overline{R}^s_{\mu\mu}$ does not define uniquely the values that $P^s_{\ell\ell}$ and $S^s_{\ell\ell}$ can assume and how the SM point is compatible with state-of-the-art theoretical and experimental results. Interestingly, non-negligible  NP effects are allowed but the values of the NP parameters $P^s_{\ell\ell}$ and $S^s_{\ell\ell}$ are rather unconstrained. To obtain stronger bounds on the values of $P^s_{\ell\ell}$ and $S^s_{\ell\ell}$, more observables sensitive to these NP contributions are required. This will be the topic of the following section.

\section{The Observable $\mathcal{A}^{\ell\ell}_{\Delta \Gamma_s}$}\label{Sec:ADG}

The untagged decay rate gives us access to the observable $\mathcal{A}^{\ell\ell}_{\Delta \Gamma_s}$  sensitive to the scalar and pseudoscalar functions $P^s_{\ell\ell}$ and $S^s_{\ell\ell}$ \cite{Bsmumu-ADG}:

\begin{eqnarray}\label{untagged}
\langle \Gamma(B_s(t)\to \ell^+\ell^-)\rangle
&\propto& e^{-t/\tau_{B_s}}\bigl[\cosh(y_st/ \tau_{B_s})+   {\cal A}^{\ell\ell}_{\Delta\Gamma_s}
\sinh(y_st/ \tau_{B_s})\bigr],\nonumber
\end{eqnarray}

where

\begin{eqnarray}
{\cal A}^{\ell\ell}_{\Delta\Gamma_s} = \frac{|P^s_{\ell\ell}|^2\cos(2\varphi_{P_s}^{\ell\ell}-\phi_s^{\rm NP}) - 
	|S^s_{\ell\ell}|^2\cos(2\varphi_{S_s}^{\ell\ell}-\phi_s^{\rm NP})}{|P^s_{\ell\ell}|^2 + |S^s_{\ell\ell}|^2}.\label{Aobs}
\end{eqnarray}

\noindent
The observable ${\cal A}^{\ell\ell}_{\Delta\Gamma_s}$ obeys the model-independent bounds
$-1\leq\mathcal{A}_{\Delta\Gamma_s}^{\ell\ell}\leq+1$,
in particular within the SM we have ${\cal A}^{\ell\ell}_{\Delta\Gamma_s}|_{\rm SM}=+1$. The ``effective life-time''

\begin{eqnarray}
\tau^s_{\ell\ell} \equiv \frac{\int^\infty_0 t\,
	\langle\Gamma(B_s(t)\to \ell^+ \ell^-)\rangle\, dt}{\int_0^\infty \langle
	\Gamma(B_s(t)\to \ell^+ \ell^-)\rangle\, dt}
\end{eqnarray}
is equivalent to ${\cal A}^{\ell\ell}_{\Delta\Gamma_s}$. As a matter of fact both observables are related through

\begin{eqnarray}
{\cal A}^{\ell\ell}_{\Delta\Gamma_s}  = \frac{1}{y_s}\left[\frac{(1-y_s^2)\tau^s_{\ell\ell}-(1+
	y_s^2)\tau_{B_s}}{2\tau_{B_s}-(1-y_s^2)\tau^s_{\ell\ell}}\right].\nonumber
\end{eqnarray}

\noindent
The first determination of ${\cal A}^{\mu\mu}_{\Delta\Gamma_s}$ (associated with the decay
$B_s\rightarrow \mu^+\mu^-$) has been performed by LHCb \cite{LHCb-2017} 
\begin{eqnarray}
\tau^s_{\mu\mu} = \left[2.04 \pm 0.44 ({\rm stat}) \pm 0.05 ({\rm syst}) \right]\hbox{ps}.
\end{eqnarray}
\noindent
This result can be converted into

\begin{eqnarray}\label{eq:ADG}
\mathcal{A}_{\Delta\Gamma_s}^{\mu\mu} = 8.24 \pm 10.72,
\end{eqnarray}
\noindent
which saturates the model-independent bounds previously discussed. However, future improvements on the measurement of $\mathcal{A}_{\Delta\Gamma_s}^{\mu\mu}$ will allow us to obtain stronger constraints on $P^s_{\mu\mu}$ and $S^s_{\mu\mu}$. For instance, setting  ${\cal A}^{\ell\ell}_{\Delta\Gamma_s}=-1, 0, 1$ singles out straight lines in Fig. \ref{fig:PvsSoAGF}

\begin{figure}
	\begin{center}
		\label{fig:flowchart}
		%\vspace{1.6cm}
		\frame{\includegraphics[width=0.8\textwidth]{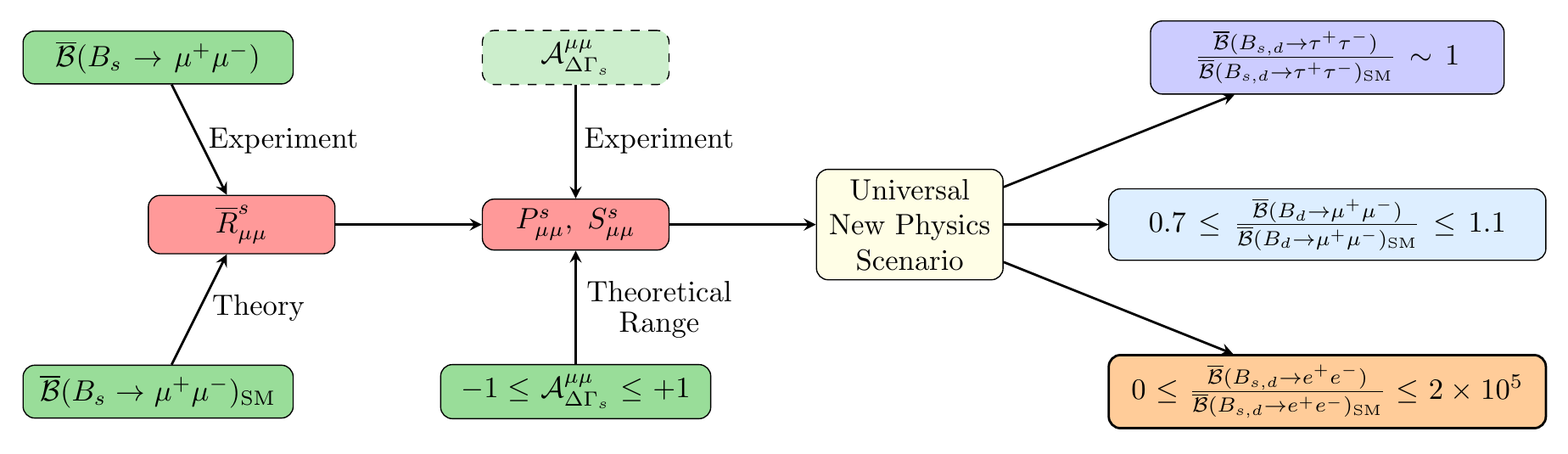}}		
	\end{center}
	\caption{Strategy to map out possible CP-violating NP contributions to $B_s\rightarrow \mu^+ \mu^-$ into other rare $B$ decay observables.}
\end{figure}

\section{Impact of $B_s\rightarrow \mu^+ \mu^-$ on other rare decays}

Working under the assumption of real $P^s_{\mu\mu}$ and $S^s_{\mu\mu}$, we will explore the effects of the bounds established from 
the  experimental results in Eqs. (\ref{eq:Rres}) and (\ref{eq:ADG})
on other rare decays. To achieve this target, we have to make assumptions to correlate the Wilson coefficients $C^{\mu \mu}_{S, P}$ for $B_{s}\rightarrow \mu^+ \mu^-$ with those for the transitions 
$B_{s}\rightarrow e^+e^-, \tau^+\tau^-$ and $B_{d}\rightarrow e^+e^-, \mu^+ \mu^-,  \tau^+\tau^-$.  In particular we will explore the implications of universal Wilson coefficients: $C^{\mu\mu(')}_{S, P}=C^{\tau\tau(')}_{S, P}=C^{e e(')}_{S, P}$ and refer to this model as ``Universal New Physics Scenario'' (UNPS). The full strategy to be followed is summarized in the flow chart in Fig. \ref{fig:flowchart}.\\

\subsection{New Physics in  $B_d\rightarrow \mu^+ \mu^-$ }
Let us first discuss potential NP effects on $B_{d}\rightarrow \mu^+ \mu^-$. We start by considering the ratio

\begin{eqnarray}
U^{ds}_{\mu\mu}\equiv \sqrt{\frac{|P^d_{\mu\mu}|^2+|S^d_{\mu\mu}|^2}{|P^s_{\mu\mu}|^2+|S^s_{\mu\mu}|^2}}
\propto \Bigl(\frac{f_{B_s}}{f_{B_d}}\Bigl)^2\Bigl|\frac{V_{ts}}{V_{td}}\Bigl|^2
\frac{\overline{\mathcal{B}}(B^0_d\rightarrow \mu^+\mu^-)}{\overline{\mathcal{B}}(B^0_s\rightarrow \mu^+\mu^-)}.
\end{eqnarray}
\noindent
In the SM, this observable assumes the value $U^{ds}_{\mu\mu}|_{\rm SM}=1$. However, possible NP effects can induce important deviations. Current data give $U^{ds}_{\mu\mu}=1.26\pm 0.49$, in good agreement with the SM.
Let us now briefly discuss
the consequences of the UNPS where the Wilson coefficients associated with the transitions  $b\rightarrow d \mu^+\mu^-$ and $b\rightarrow s \mu^+\mu^-$ turn out to be the same. As shown in \cite{Fleischer:2017ltw} the final result is a linear correlation between $\overline{\mathcal{B}}(B_s\rightarrow \mu^+ \mu^-)$ and $\overline{\mathcal{B}}(B_d\rightarrow \mu^+ \mu^-)$. To quantify the possible NP contributions we evaluate the ratio in Eq.~(\ref{Rellells}) adapted for $b \rightarrow d$ transitions yielding 

\begin{eqnarray}
0.65 \leq \overline{R}_{\mu\mu}^d \leq 1.11.
\end{eqnarray}

\subsection{New Physics in  $B_{d, s}\rightarrow \tau^+ \tau^-$ and $B_{d, s}\rightarrow e^+ e^-$}

To study the implications of the UNPS on the decays $B_{s, d}\rightarrow \tau^+ \tau^-,~e^+ e^-$ we start by writing the following equations that establish the relationship between $P^s_{\mu\mu}$, $S^s_{\mu\mu}$ and 
the corresponding functions for generic leptons in the final state

\begin{eqnarray}\label{eq:conversion}
P^s_{\ell\ell} = \left(1-\frac{m_\mu}{m_\ell}\right){\cal C}_{10}+ \frac{m_\mu}{m_\ell}P^s_{\mu\mu},\quad\quad\quad
S^s_{\ell\ell} = \frac{m_\mu}{m_\ell}
\sqrt{\frac{1-4\frac{m_\ell^2}{M_{B_s}^2}}{1-4\frac{m_\mu^2}{M_{B_s}^2}}}S^s_{\mu\mu}.
\end{eqnarray}
\noindent
To assess the implications of the UNPS on $\mathcal{B}(B_s\rightarrow \tau^+ \tau^-)$ we substitute $m_{\ell}=m_{\tau}$ in Eq.~(\ref{eq:conversion}). The resulting ratio 

\begin{eqnarray}\label{eq:ratiotau}
\frac{m_{\mu}}{m_{\tau}}=0.059 
\end{eqnarray}

\noindent
in front of the potential NP contributions in $P^s_{\mu\mu}$ and $S^s_{\mu\mu}$ on the right-hand side of Eq.~(\ref{eq:conversion}) acts as a suppression factor. Therefore, within the UNPS the branching fraction $\mathcal{B}(B_s\rightarrow \tau^+ \tau^-)$ experiences mild deviations with respect to the SM prediction.
The ratio introduced in Eq.~(\ref{Rellells}) leads to

\begin{eqnarray}
0.8\leq \overline{R}^s_{\tau\tau} \leq 1.0.
\end{eqnarray}
\noindent
Additionally, the UNPS allows us to predict 
\begin{eqnarray}
0.995\leq \mathcal{A}_{\Delta\Gamma_s}^{\mu\mu}\leq 1.000,
\end{eqnarray}
where again there is a tiny deviation with respect to the SM value. The effects on the corresponding observables for $B_d\rightarrow \tau^+ \tau^-$ are similar and can be found in \cite{Fleischer:2017ltw}.\\

\noindent
So far, the predictions of the UNPS have been consistent with the expectations from the SM. However, for the decays $B_{s,d}\rightarrow e^+ e^-$ the effect can potentially be dramatically different. To understand this result consider that for $\ell=e$  the functions $P^s_{\mu\mu}$ and $S^s_{\mu\mu}$ are mapped out into $P^s_{ee}$ and $S^s_{ee}$ through the ratio
\begin{eqnarray}\label{eq:ratioel}
\frac{m_{\mu}}{m_{e}}=206.77.
\end{eqnarray}

\noindent
Unlike the case for the $\tau$ leptons in Eq.~(\ref{eq:ratiotau}), 
our ``conversion factor'' in Eq.~(\ref{eq:ratioel}) enhances the potential NP contributions inside $P^s_{\mu\mu}$ and $S^s_{\mu\mu}$. Thus, instead of acting as a suppressor, the mass of the electron works as an enhancement factor. Within the UNPS we can make the following predictions for $\mathcal{B}(B_s\rightarrow e^+ e^-)$: 

\begin{eqnarray}\label{eq:e_enhancement}
0\leq 
\overline{R}^s_{ee}
\leq 1.7\times 10^{5},\quad\quad 0\leq \frac{\overline{\mathcal{B}}(B_s \rightarrow e^+e^-)}{\overline{\mathcal{B}}(B_s \rightarrow \mu^+\mu^-) } \leq 4.8.
\end{eqnarray}

\noindent
Surprisingly the upper bound of the previous inequalities lies just a factor of 20 below the limit that CDF determined in 2009. In view of these results, we encourage the experimental search of $B_s \rightarrow e^+e^-$, since any measurement of $\overline{\mathcal{B}}(B_s \rightarrow e^+ e^-)$ within the capabilities of current or future experiments would be an unambiguous signal of NP.~For $B_d \rightarrow e^+e^-$ the enhancement is such that the branching fraction is just one order of magnitude below the results presented in Eq.~(\ref{eq:e_enhancement}). The pattern of predictions discussed requires that the effective couplings (Wilson coefficients) are independent of the mass of the lepton in the final state, this is a very restrictive condition that does not materialize in scenarios such as the Minimal Supersymmetric SM.

\section{Impact of New Sources of CP violation}

Until now only trivial values for the CP violating phases $\varphi^{\mu\mu}_P$ and $\varphi^{\mu\mu}_S$  have been assumed. In this section we relax this condition and discuss the impact of non-vanishing phases on $B_s\rightarrow  \mu^+\mu^-$.\\

\noindent
Firstly,  we use the following time-dependent asymmetry 

\begin{equation}
\frac{\Gamma(B^0_s(t)\to \mu_\lambda^+\mu^-_\lambda)-
	\Gamma(\bar B^0_s(t)\to \mu_\lambda^+
	\mu^-_\lambda)}{\Gamma(B^0_s(t)\to \mu_\lambda^+\mu^-_\lambda)+
	\Gamma(\bar B^0_s(t)\to \mu_\lambda^+\mu^-_\lambda)}
=\frac{ {\cal C}_{\mu\mu}^\lambda  \cos(\Delta M_st)+{\cal S}_{\mu\mu}^\lambda
	\sin(\Delta M_st)}{\cosh(y_st/\tau_{B_s}) + 
	{\cal A}_{\Delta\Gamma_s}^{\lambda, \mu\mu} \sinh(y_st/\tau_{B_s})},\nonumber
\end{equation}
to introduce the observables \cite{Bsmumu-ADG,BFGK}: ${\cal C}_{\mu\mu}^\lambda$ and ${\cal S}_{\mu\mu}^\lambda$. They are given by

\begin{eqnarray}
{\cal C}_{\mu\mu}^\lambda &= & -\eta_\lambda\left[\frac{2|P^s_{\mu\mu}S^s_{\mu\mu}|\cos(\varphi^{\mu\mu}_P-
	\varphi^{\mu\mu}_S)}{|P|^2+|S|^2} 
\right] \equiv -\eta_\lambda{\cal C}_{\rm \mu\mu},\label{Cobs}\\
{\cal S}_{\mu\mu}^\lambda
&=&\frac{|P^s_{\mu\mu}|^2\sin(2\varphi^{\mu\mu}_P-\phi_s^{\rm NP})-|S^s_{\mu\mu}|^2\sin(2\varphi^{\mu\mu}_S-\phi_s^{\rm NP})}{|P^s_{\mu\mu}|^2+|S^s_{\mu\mu}|^2}
\equiv {\cal S}_{\mu\mu}\label{Sobs},
\end{eqnarray}
where $\eta_{L/R}=\pm 1$ defines the helicity of the final state $\mu$ leptons. Within the SM, ${\cal C}_{\rm \mu\mu}={\cal S}_{\mu\mu}=0$.  
Notice that the observables $\mathcal{A}^{\lambda, \mu\mu}_{\Delta\Gamma_s}$, ${\cal C}_{\mu\mu}$ and ${\cal S}_{\mu\mu}$ do not depend on the helicity of the final state leptons. Moreover they are extremely clean since they are free from hadronic parameters. The three CP asymmetries are not independent because they fulfil the condition 
\begin{eqnarray}
({\cal A}^{\mu\mu}_{\Delta\Gamma_s})^2 + ({\cal S}_{\mu\mu})^2 + ({\cal C}_{\mu\mu})^2=1.
\end{eqnarray}
\noindent
Unfortunately, in general $\mathcal{A}^{\mu\mu}_{\Delta\Gamma_s}$, ${\cal C}_{\mu\mu}$, ${\cal S}_{\mu\mu}$ and $\bar{R}^s_{\mu\mu}$ do not provide sufficient information to fully determine the magnitudes and phases of all the complex coefficients $C^{\mu\mu (')}_S$ and $C^{\mu\mu (')}_P$ inside  $P^s_{\mu\mu}$ and $S^s_{\mu\mu}$. This is only possible within specific frameworks where extra assumptions reduce the number of unknowns.
Here we consider the SMEFT \cite{AGMC}, where the following conditions hold: $C_P^{\mu\mu}=-C_S^{\mu\mu}$, $C_P^{\mu\mu'}=C_S^{\mu\mu'}$. Moreover, we explore two cases found frequently in the literature: $C^{\mu\mu}_{S}=0$ and $C_{S}^{\mu\mu'}=0$. If we introduce the parameter $x=C_{S}^{\mu\mu'}/C^{\mu\mu}_{S}$, then these situations correspond to $x\rightarrow \infty$ and $x=0$, respectively. It is then possible to show that only two independent parameters are required we choose them to be $|S^s_{\mu\mu}|$ and $\varphi^{\mu\mu}_S$. The full set of Wilson coefficients 
can then be related to $|S^s_{\mu\mu}|$ and $\varphi^{\mu\mu}_S$ as explained in  \cite{Fleischer:2017yox}.\\

\begin{figure}\label{fig:regions}
	\includegraphics[width=0.45\textwidth]{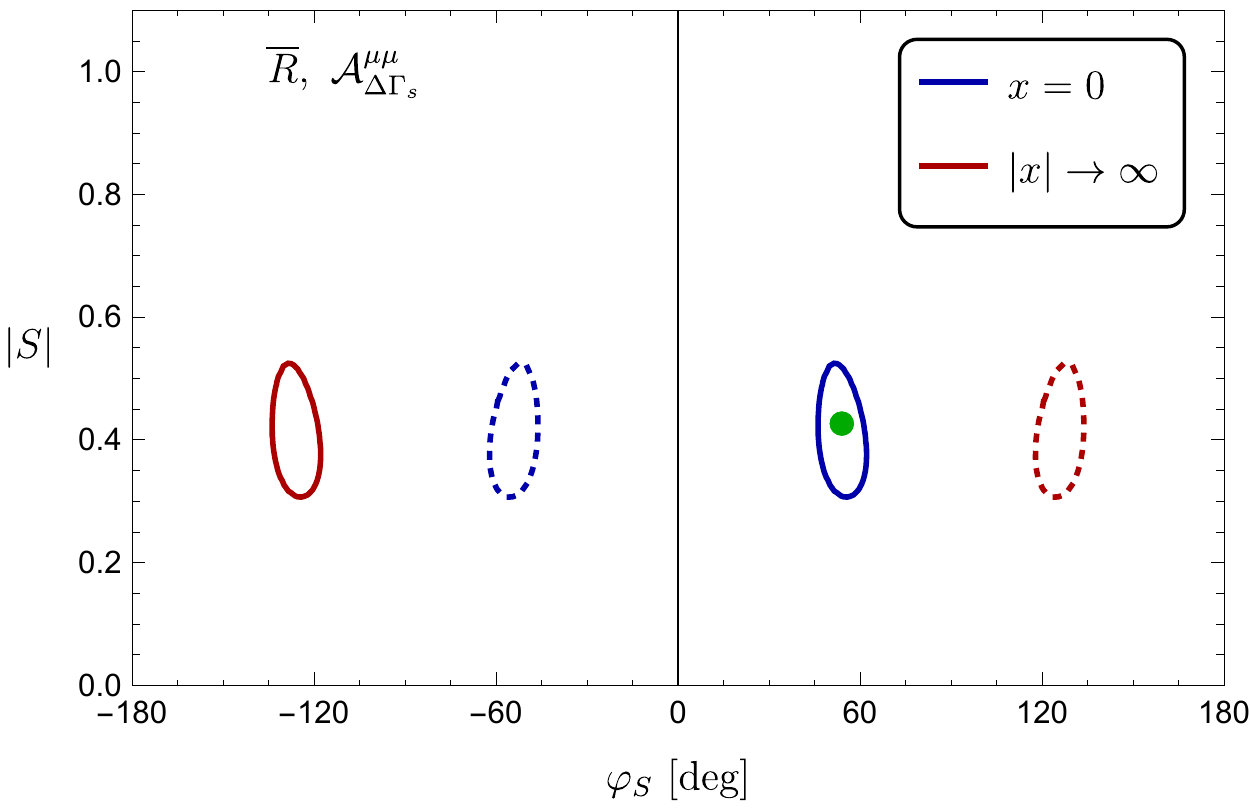}
	\includegraphics[width=0.45\textwidth]{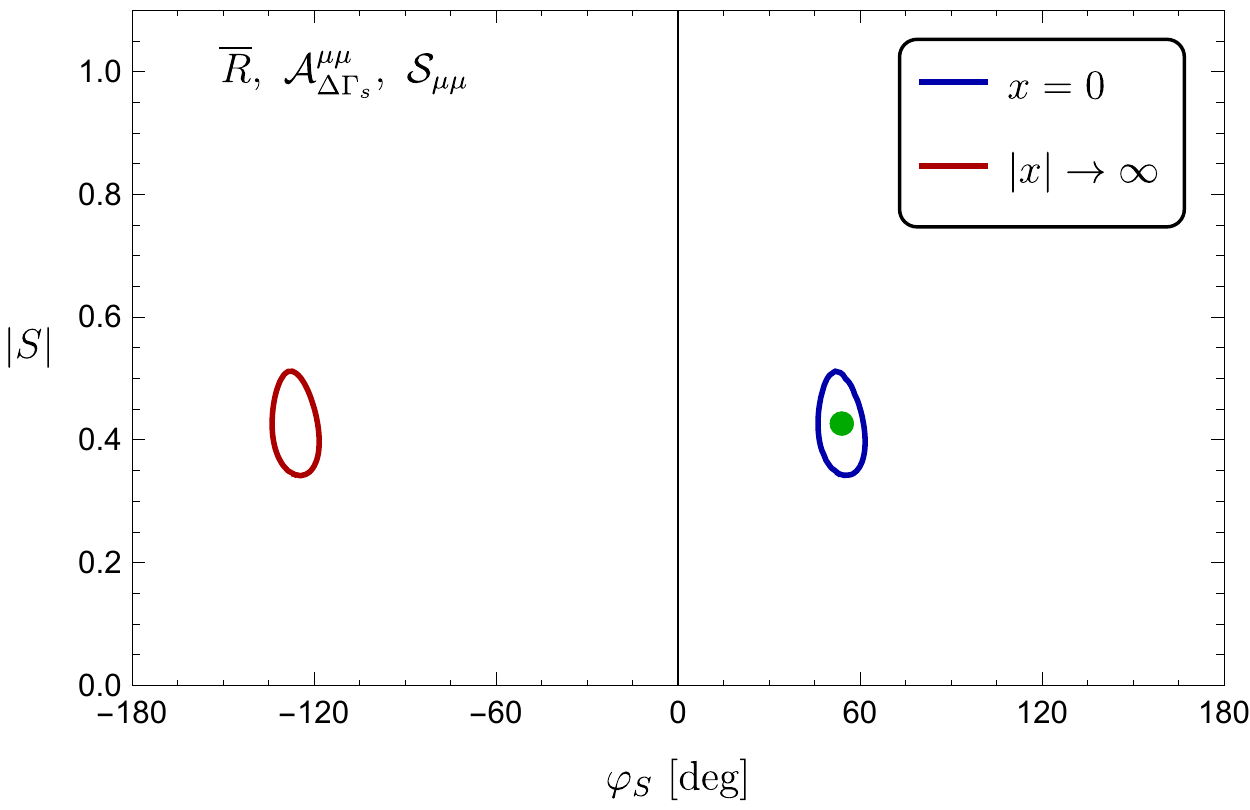}
	\caption{$\chi^2$ fits used in the determination of the allowed regions compatible with the SMEFT model: $x=0$ and $x=\infty$ corresponding to $C_S^{\mu\mu'}=0$ and  $C_S^{\mu\mu}=0$ respectively.}
	% 	\vspace{0.3cm}
	% 	\textcolor[rgb]{0,0,1}{Non-zero value for $|S|$ at $5~\sigma$ level}
\end{figure}

\noindent
We illustrate our strategy with an example, considering the following set of hypothetical measurements:

\begin{equation} \label{eq:obsx0fit}
{\cal A}_{\Delta\Gamma_s}^{\mu\mu} = 0.58 \pm 0.20, 
\quad {\cal S}_{\mu\mu} = -0.80 \pm 0.20, \quad {\cal C}_{\mu\mu} = 0.16 \pm 0.20.\nonumber
\end{equation}
\\
\noindent
We perform $\chi^2$-fits to our assumptions $x\rightarrow \infty$ and $x=0$. Then, we profile over the two independent parameters  to obtain the regions shown in Fig. \ref{fig:regions}. On the left side of the figure, the observables $\bar{R}^s_{\mu\mu}$ and $\mathcal{A}^{\mu\mu}_{\Delta \Gamma}$ single out fours regions compatible with our hypotheses. If, in addition, we include $S_{\mu\mu}$, we can eliminate the two subregions enclosed inside the dashed contours leading to the plot on the right.  Here both scenarios ($x=0$ and $x=\infty$) are still possible. However the sign of ${\cal C}_{\mu\mu}$ solves this ambiguity since in our example we have  $0<{\cal C}_{\mu\mu}$ for $x=0$ whereas ${\cal C}_{\mu\mu}<0$ for $x\rightarrow\infty$.

%\vspace{0.2cm}
\section{Outlook}
In spite of being strongly suppressed in the SM, the decay processes $B_{s, d}\rightarrow \ell^+\ell^-$  have the potential to open new avenues in our quest for NP effects and offer a rich phenomenological structure. The measurement of the branching fraction  $\overline{\mathcal{B}}(B_s\rightarrow \mu^+\mu^-)$ has been found consistent with the SM prediction. However, it leaves plenty of room for new scalar and pseudoscalar interactions. The observable ${\cal A}_{\Delta\Gamma_s}^{\mu\mu}$ arises once neutral $B_s$ mixing is taken into account and is a powerful tool for solving ambiguities between different NP scenarios. Under the assumption of universal short-distance contributions, we have mapped out the current constraints on the observables for $B_s\rightarrow \mu^+\mu^-$ into the corresponding ones for $B_{s, d}\rightarrow \tau^+ \tau^-$ and  $B_{s, d}\rightarrow e^+e^-$ showing how in the first case possible NP contributions are suppressed by a factor $1/m_{\tau}$.  In contrast, in the second case they can be dramatically enhanced by the ratio $1/m_{e}$, having the potential of becoming accessible within the realm of current and foreseeable experiments. Hence their search is strongly encouraged. Finally, in addition to the branching fractions, rare $B$ decays provide extra observables sensitive to  CP-violating NP phases that can be extremely valuable for unveiling new sources of CP violation.

\section{Acknowledgements}

I would like to thank Robert Fleischer, Ruben Jaarsma and Daniela Galarraga-Espinosa for a very fruitful and enjoyable collaboration. The research projects discussed here have been supported by the Netherlands Foundation for Fundamental Research
of Matter (FOM) programme 156,  ``Higgs as Probe and Portal'', and by the
National Organisation for Scientific Research (NWO).

\end{document}